# Cybernetic Principles of Aging and Rejuvenation:

# the buffering-challenging strategy for life extension

Francis Heylighen

*Evolution, Complexity and Cognition group*
*Vrije Universiteit Brussel*

**Abstract**: Aging is analyzed as the spontaneous loss of adaptivity and increase in fragility that characterizes dynamic systems. Cybernetics defines the general regulatory mechanisms that a system can use to prevent or repair the damage produced by disturbances. According to the law of requisite variety, disturbances can be held in check by maximizing buffering capacity, range of compensatory actions, and knowledge about which action to apply to which disturbance. This suggests a general strategy for rejuvenating the organism by increasing its capabilities of adaptation. Buffering can be optimized by providing sufficient rest together with plenty of nutrients: amino acids, antioxidants, methyl donors, vitamins, minerals, etc. Knowledge and the range of action can be extended by subjecting the organism to an as large as possible variety of challenges. These challenges are ideally brief so as not to deplete resources and produce irreversible damage. However, they should be sufficiently intense and unpredictable to induce an overshoot in the mobilization of resources for damage repair, and to stimulate the organism to build stronger capabilities for tackling future challenges. This allows them to override the trade-offs and limitations that evolution has built into the organism's repair processes in order to conserve potentially scarce resources. Such acute, "hormetic" stressors strengthen the organism in part via the "order from noise" mechanism that destroys dysfunctional structures by subjecting them to strong, random variations. They include heat and cold, physical exertion, exposure, stretching, vibration, fasting, food toxins, micro-organisms, environmental enrichment and psychological challenges. The proposed buffering-challenging strategy may be able to extend life indefinitely, by forcing a periodic rebuilding and extension of capabilities, while using the Internet as an endless source of new knowledge about how to deal with disturbances.

**Introduction**

Aging is commonly understood as the gradual deterioration of the organism due to the passage of time. However, it is surprisingly difficult to establish which processes of deterioration are caused by the inescapable passage of time, and which merely by avoidable, external causes, such as diseases, injuries, or an unhealthy diet. Common symptoms of aging, such as the loss of bone mass, muscular strength and joint flexibility, often turn out to be not just preventable, but reversible, given the correct regime of activity, diet or supplements. Therefore, aging is defined in biology not as deterioration of specific functions or systems, but as a general increase in the rate of mortality with age (Mueller, Rauser, & Rose, 2011): once you have passed your "essential lifespan" (Rattan, 2012), the probability that you would die in the next year increases with every year that passes—until, around the age of 120, it reaches 100%.

The cause of this increase can be conceived as growing *fragility*: the body becomes increasingly less likely to recover from some disturbance, such as an infection, a fall, or some stress on a particular organ or vessel. In that sense, people do not die from "old age": they always die from a specific cause, such as an accident, a stroke, a tumor, or a case of pneumonia. Age is merely an indicator of a decrease in robustness, resilience or the ability to recover from such problems. The lower the body's resilience, the higher the probability that it will not survive a given disturbance. Therefore, aging could be redefined as the increase of fragility with the passage of time. Rejuvenation can then be defined as a process that increases resilience, or reduces fragility.

The first question we must ask is why fragility tends to increase with time. In other words, why do we age? The second question is then in how far we can slow down or reverse such loss of resilience. In other words, how can we extend life, or even rejuvenate the organism?

Both questions can be investigated from the perspective of *cybernetics*, the science of adaptive, goal-directed systems (Ashby, 1964; Heylighen & Joslyn, 2003). All organisms are cybernetic systems: they try to reach their goals by counteracting any disturbances that make them deviate from this goal state. This is the process of *regulation* or *control*. The most basic goal of a living system is simply survival. This means that it needs to evade or suppress any disturbance that threatens to kill it: a predator, a cold spell, an injury, an infection, a tumor, a lack of nutrients, ... The degree to which an organism is capable of successfully counteracting such a variety of disturbances determines its adaptivity or resilience, and thus its probability of survival.

Cybernetics studies not just this process of goal-directed action that characterizes living systems. It also investigates the more primitive mechanism of self-organization that characterizes any dynamic system (Ashby, 1962). This process, by which a system spontaneously evolves towards a reduced number of states, may explain why fragility increases in the absence of any outside intervention. The mechanism of control may then explain what is needed to counteract or even reverse this descent into a rigidity.

The following sections will first review the two cybernetic mechanisms, while considering their implications for aging and rejuvenation. I will then propose a number of concrete measures inspired by these cybernetic principles that are likely to stave off aging and boost general health, robustness and fitness, and this without apparent limit. These measures can be subdivided in two main categories, a conservative or defensive strategy that I will call *buffering*, and that aims at protecting the organism against damage, and a progressive or offensive strategy, called *challenging*, that aims at rebuilding already damaged structures and at increasing the organism's capabilities for counteracting further sources of damage.

## Aging as a descent into rigidity

Ashby (Ashby, 1962), one of the founders of cybernetics, analyzed self-organization as the automatic process by which a dynamic system moves towards a state of equilibrium, or what we would nowadays call an *attractor*. A dynamic system is a system that undergoes changes determined by the intrinsic forces or "dynamics" that govern its evolution. These changes are modeled as movements along a trajectory that meanders through the system's *state space*—i.e. the set of all the states or configurations that the system could possibly be in. Such changes are typically irreversible: while the system can move from state *A* (say, young and vigorous) to state *B* (say, old and decrepit), it in general cannot return from *B* to *A*. Even when the changes are reversible in principle, the ease or probability of a move in one direction (say, $A \rightarrow B$) will in general be different from a move in the opposite direction ($B \rightarrow A$) (Heylighen, 1992).

The result of this time asymmetry is that the system will tend to leave the states from which it can move on easily (say, *A*), while settling in the states from which it is difficult to leave (say, *B*). A region in state space that the system can enter but not leave is called an *attractor*. An attractor is surrounded by an attractor *basin*: these are the states that the system tends to leave in order to end up in the attractor, but to which it cannot return afterwards (Heylighen, 2001). Once in the attractor, the system by definition has to stay there. This means that from now on the system's further evolution is strictly limited or constrained: it has lost the freedom of visiting states outside the attractor; the system has become more rigid.

The simplest attractor is a point attractor, which consists of a single equilibrium state. Once the system has reached this state, it is no longer able to move to any other state. The only true equilibrium state for a living organism is death. For more complex dynamic systems, attractors tend to have more complicated shapes. This means that movement is still possible within the attractor, but escape from the attractor remains foreclosed.

Truly complex systems, however, are not deterministic but probabilistic or *stochastic*: the system can move from a given state *A* to several other states *B*, *C*, *D*, … each with a certain transition probability. This unpredictability results from the fact that there are always unknown factors or disturbances that make the system move in unforeseen directions. For such stochastic systems, there are in general no strict attractors, since external events can always push the system out of an attractor. However, we can still distinguish "approximate" attractors. These are regimes of activity that the system is unlikely to leave, although exceptional circumstances may still push it out of there.

In this case, the strict separation attractor-basin needs to be replaced by a more fuzzy degree of "attractiveness" characterizing a state (or region) *A*. This can be defined mathematically as the probability that the system would enter *A* minus the probability that it would leave *A* (Gershenson & Heylighen, 2003). Attractiveness can be visualized as a combination of the "width" and "depth" of the attractor valley: the wider the surrounding basin, the easier it is to get into the valley and from there to slide into to attractor at the bottom; the deeper the valley, the more difficult it is to get out…

This abstract representation of time-asymmetric changes provides us with a simple model of aging. As an organism undergoes changes under the influences of various internal and external forces, it tends to preferentially move toward an (approximate) attractor. This means that these changes become increasingly more difficult to undo, while the range of potential further changes becomes increasingly limited. Thus, the system's range of possible reactions to different challenges becomes ever smaller. Rattan (Rattan, 2007, 2008, 2012) formulates this process as a reduction of the "homeodynamic space". The homeodynamic space is that part of the system's state space in which it should be able to move in order to adapt to the disturbances that may push it out of its viable region. As that space shrinks, the system becomes increasingly rigid, and thus brittle. The final result is an attractor state from which escape is no longer possible: death.

Note that the second law of thermodynamics—according to which closed systems move towards an equilibrium state in which entropy is maximized—is merely a special case of this general dynamics. The second law does not strictly apply to living systems because these are open, and thus able to get rid of entropy by dissipating it into their environment. Therefore, the second law on its own cannot be used to explain aging. The second law also assumes that the final equilibrium is characterized by maximal disorder or dissipation. This is the natural outcome for a system whose dynamics is time-symmetric at the microscopic level, but irreversible at the macroscopic, thermodynamic level (Gershenson & Heylighen, 2003).

However, in general, systems are time-asymmetric at all levels. This allows for a final attractor that is ordered (e.g. a crystalline structure) rather than disordered (e.g. a homogeneous gas). That explains why Ashby used this dynamics to explain self-organization, rather than the disorganization that is typical of entropy increase. But "organization" in Ashby's sense refers to a structure that is more constrained or rigid than the one that came before. Therefore, self-organization is not necessarily a good thing: that depends on your goals and perspective (Gershenson & Heylighen, 2003). The death and fossilization of an organism can be seen as the self-organization of a structured piece of rock!

The irreversibility of such dynamics, whether it ends up in order or disorder, implies a loss of information. A wide variety of different states end up in the same attractor, thus losing the characteristics that differentiated them. Entropy is equivalent to uncertainty, i.e. absence of information (Heylighen & Joslyn, 2003; Heylighen, 2001). But even a point attractor—which has zero entropy—while providing certainty about the present state of the system, has lost the information about the initial state from which the system started. In other words, reaching an attractor erases the memory about what happened to the system before it entered the attractor.

The conclusion is that any dynamic system, including living organisms, has a tendency to evolve towards an increasingly rigid state over time, while "forgetting" any information that used to be stored in its earlier states. However, in open systems subject to outside influences such evolution is not predetermined, because disturbances can push it out of its attractor. Let us then investigate the processes that may prevent or reverse this loss of adaptivity and memory.

## Regulation and the law of requisite variety

The most fundamental mechanism studied in cybernetics is *regulation* or *control*, in which a system tries to reach a particular goal state or goal region in the face of internal or external disturbances that make it deviate from that goal. Effective control combines three fundamental mechanisms (Heylighen & Joslyn, 2003): *feedforward*, *feedback* and *buffering*.

In feedforward regulation, the system acts on the disturbance *before* it has affected the system, in order to prevent a deviation from happening. For example, if you perceive a sudden movement in the vicinity of your face, you will close your eyelids before any projectile can hit it, so as to prevent potential damage to your eyeball. The disadvantage that it is not always possible to act in time, and that the anticipation may turn out to be incorrect, so that the action does not have the desired result. For example, the projectile may not have been directed at your eyes, but at a different part of your face. By shutting your eyes, you make it more difficult to avoid the actual impact.

In feedback regulation, the system neutralizes or compensates the deviation *after* the disturbance has pushed the system away from its goal, by performing an appropriate repair action. For example, a thermostat compensates for a fall in temperature by switching on the heating, but only after it detected a lower than desired temperature. For effective regulation, it suffices that the feedback is negative—i.e. reducing the deviation—because a sustained sequence of corrections will eventually suppress any deviation. The advantage is that there is no need to rely on a complex, error-prone process of anticipation on the basis of imperfect perceptions: only the direction of the actual deviation has to be sensed. The disadvantage is that the counteraction may come too late, allowing the deviation to cause irreversible damage before it was effectively suppressed.

Buffering—at least in the cybernetic sense—is a *passive* form of regulation: the system dampens or absorbs the disturbances through its sheer bulk of protective material. Examples of buffers are shock absorbers in a car, water reservoirs or holding basins dampening fluctuations in rainfall, and the fur that protects a warm-blooded animal from variations in outside temperature. The advantage is that no energy, knowledge or information is needed for active intervention. The disadvantage is that buffering is not sufficient for reaching goals that are not equilibrium states in themselves, because moving away from equilibrium requires active intervention. For example, while a reservoir makes the flow of water more even, it cannot provide water in regions where it never rains. Similarly, fur alone cannot maintain a mammal body at a temperature higher than the average temperature of the surroundings: that requires active heat production.

Active regulation (feedforward and feedback) is subject to Ashby's *law of requisite variety* (Ashby, 1958, 1964): the greater the variety of disturbances the system may be confronted with, the greater the variety of actions the system should be able to perform in order to prevent or compensate the corresponding deviations from its goal state. Moreover, it is subjected to what has been called the law of regulatory models (Conant & Ashby, 1970) or the *law of requisite knowledge* (Heylighen & Joslyn, 2003; Heylighen, 1992): for each disturbance (or deviation) encountered, the system should know which of its variety of actions is most appropriate to compensate that particular disturbance. Passive regulation (buffering) is indifferent as to the specific action needed, because it automatically dampens all disturbances of a particular type. These three requirements for successful regulation are summarized in the *extended form of the law of requisite variety* (Aulin-Ahmavaara, 1979; Heylighen & Joslyn, 2003), which is expressed by the following inequality:

$$H(E) \geq H(D) - H(A) + H(A|D) - B \tag{1}$$

*H* here represents the Shannon uncertainty (statistical entropy) of a probability distribution. This is a mathematical measure for the range of possible outcomes: the larger *H*, the wider the range of possibilities, and the more uncertain we are about which possibility will be realized. *E* represents the *essential variables* of the cybernetic system (Ashby, 1964), which represent features of the system that must be protected for the system to survive. Examples of essential variables for the human body are temperature, blood pressure, plasma glucose level, and oxygen level. These must be kept within a close range around their ideal value (goal). Movement outside of that range is likely to produce irreparable damage. For example, a lack of oxygen (e.g. caused by an obstruction of blood flow during a stroke) may damage the brain, heart or other vital organs.

This means that $H(E)$ should preferably be kept as small as possible. In other words, any deviations from the ideal values must be efficiently suppressed by the control mechanism. The inequality expresses a lower bound for $H(E)$: it cannot be smaller than the sum on the right-hand side. That means that if we want to make $H(E)$ smaller, we must try to make the right-hand side of the inequality smaller. This side consists of four terms, expressing respectively the variety of disturbances $H(D)$, the variety of compensatory actions $H(A)$, the lack of requisite knowledge $H(A|D)$ and the buffering capability *B*. Let us discuss each of these terms in more detail.

Obviously, the range of outcomes for the essential variables $H(E)$ will be larger for a larger range of disturbances $H(D)$ affecting these variables. This means that organisms in complex, variable environments run more risk of encountering some disturbance that will drive their essential variables outside of their safety range $H(E)$. That suggests the first, most obvious strategy for safeguarding life: avoiding environments or situations in which a wide variety of dangers may occur. You are less likely to die while sitting in your armchair than while trekking through the Amazonian rain forest or wandering through the concrete jungle of a megacity.

The second term $H(A)$ (requisite variety of actions) expresses your capacity to compensate for such disturbances. The larger the variety of counteractions you are capable of performing, the more likely that at least one of them will be able to solve the problem, escape the danger, or restore you to a safe, healthy state. Thus, the Amazonian jungle may not be so dangerous for an explorer having a gun to shoot dangerous animals, medicines to treat disease or snakebite, filters to purify water, and the physical condition to run fast or climb in trees if threatened. The term $H(A)$ enters the inequality with a minus (–) sign, because a wider range of actions allows you to maintain a smaller range of deviations in the essential variables $H(E)$.

The third term $H(A|D)$ (lack of requisite knowledge) reminds us that action alone is not sufficient: if you do not know which action is appropriate for the given disturbance, you can only try out actions at random, in the hope that one of them will be effective (and that none of them would make your situation worse). For example, there is little use in having a variety of antidotes for different types of snakebites if you do not know which snake bit you. $H(A|D)$ expresses your uncertainty about performing an action *A* (e.g. taking a particular antidote) for a given disturbance *D* (e.g. being bitten by a particular snake). The larger your uncertainty, the larger the probability that you would choose a wrong action, and thus fail to reduce the deviation $H(E)$. Therefore, this term has a "+" sign in the inequality: more uncertainty (= less knowledge) produces more potentially lethal variation in your essential variables.

The final term *B* (buffering) expresses your amount of protective reserves or buffering capacity. Better even than applying the right antidote after a snake bite is to wear protective clothing thick enough to stop any snake poison from entering your blood stream. The term is negative because higher capacity means less deviation in the essential variables.

**Aging caused by trade-offs in damage repair**

The law of requisite variety expresses in an abstract form what is needed for an organism to prevent or repair the damage caused by disturbances. If this regulation is insufficient, damage will accumulate, including damage to the regulation mechanisms themselves. This produces an acceleration in the accumulation of damage, because more damage implies less prevention or repair of further damage, and therefore a higher rate of additional damage. This positive feedback of damage producing more damage may explain the exponential growth in mortality rates after middle age, as quantified by the classic Gompertz law (Mueller et al., 2011; Olshansky & Carnes, 1997).

The first question, though, is why the regulation mechanism would allow cumulative damage. The essence of regulation is that disturbances happen all the time, but that their effects are neutralized before they have irreparably damaged the organism. Indeed, natural selection has programmed organisms for survival, and this requires an effective cybernetic control mechanism. On the other hand, no control mechanism is powerful enough to suppress *any* potential disturbance. Therefore, in nature most organisms die from the encounter with an unusually strong disturbance— such as a predator attack, starvation, freezing, or drowning—that their regulatory mechanisms are not capable of compensating. They very rarely reach "old age", i.e. undergo the slow, but accelerating, accumulation of small amounts of functional damage that we have characterized as increasing fragility.

Therefore, natural selection for better control mechanisms has little effect on these long-term aging processes. The "force of natural selection" decreases with advancing age (Mueller et al., 2011): there is little benefit in being able to solve problems of deterioration that are unlikely to ever be encountered because the organism has already died from other causes. Hence, variations that would strengthen that capability are unlikely to be selected. Because of this weak selective pressure, our genes have not really had the chance to evolve repair mechanisms that would be useful particularly in old age.

More generally, variations that improve the organism's capacity to repair short-term damage will be preferred by natural selection over variations that increase control over long-term damage, because their effect on fitness is much more direct. When there is a competition or trade-off between short-term and long-term capabilities (e.g. because they require the same resources), the bulk of the resources will go to tackling the urgent problems.

This is the logic underlying the "triage" theory of aging (Ames, 2006). It proposes that the body allocates nutrients in priority for preventing damage with short-term consequences, thus withholding resources from repair processes whose absence would only produce real damage in the long term, i.e. in old age. For example, if combating an acute infection requires the mobilization of all the vitamin

C in the body, then processes that use vitamin C for preventing free radical damage to DNA will be temporarily interrupted. The reason is that diffuse cases of DNA damage cannot immediately kill the organism—unlike an acute infection—even though in the longer term they may lead to cancer.

Another model of aging based on such a trade-off is the "disposable soma theory" (Kirkwood & Austad, 2000; Kirkwood, 2005). It states that natural selection prioritizes successful reproduction over extending the maximum life span: once the genes are safely passed on to offspring, there is little or no fitness benefit in keeping the parent's body ("soma") alive. Therefore, deterioration tends to set in after the age of reproduction.

In conclusion, the limited availability of the resources necessary to counteract disturbances, and the fact that natural selection prefers actions with short-term fitness benefits to those with long-term benefits explains why under normal circumstances the regulation mechanisms of the body are not fully effective in preventing long-term deterioration. However, the deeper understanding of regulation provided by the law of requisite variety and of increasing fragility provided by the principle of self-organization points us towards the most fundamental factors in the aging process. These suggest concrete strategies for minimizing—or even reversing—aging.

**Anti-aging strategies**

Avoiding known dangers

The first and most obvious strategy to live long is to avoid disturbances ($H(D)$) so strong or unusual that your inbuilt regulatory system is not able to compensate for them. For example, you should avoid extended exposure to freezing temperatures, fire, poisons, potentially lethal or crippling infections, encounters with dangerous animals or people, falls from great heights, car crashes, etc. This includes disturbances that are not directly lethal, but that are known to produce accumulating damage, such as smoking, continuing pressures on joints or organs that produce wear and tear injuries, chronic inflammation, long-term exposure to toxins that tend to accumulate in the body (such as heavy metals), and diets (e.g. those with a high glycemic load, or a lot of alcohol) that gradually wear down vital organs (e.g. the liver) or functions (e.g. insulin sensitivity).

However, as we will elaborate further, this does not mean avoiding all disturbances: on the contrary, controllable disturbances actually train or fortify the control system, and thus make it less likely to succumb to unexpected disturbances.

Buffering

The buffering term $B$ in the requisite variety equation suggests the next, relatively straightforward strategy. The buffer refers to the body's reserves of matter, energy or resources that it relies on to dampen fluctuations.

For example, fat deposits and muscle tissue provide the body with insulation against cold and shocks, and reserves of calories and building blocks to be used in case of insufficient food supply. Growing lean muscle mass through exercise and a protein-rich diet is an excellent strategy for prolonging life (De Vany, 2010): it not only stores protein, but tends to be accompanied by growth in other organs that play a buffering or active role, such as bones, lungs, and heart. Fat, on the other hand, is only useful up to a point: too much fat reduces life expectancy, as fat cells release inflammatory cytokines. Moreover, the lipids stored in fat play fewer crucial roles in the organism than the amino acids that constitute proteins.

The third type of macronutrients, next to fats and proteins, are carbohydrates. These are digested into glucose that is stored as glycogen in the liver and muscles, or converted to fat. Unlike fats and proteins, carbohydrates are not essential to the body, as it can produce glucose from proteins through the mechanism of gluconeogenesis. Moreover, its capacity for storing glycogen is relatively small. Eating too large quantities of carbohydrate therefore tends to overwhelm the cells' ability to take up glucose, with the result that circulating glucose molecules damage tissues by attaching to proteins. This process of *glycation* is viewed as a key contributor to aging (Lee & Cerami, 1992; Ulrich & Cerami, 2001), and the waste materials it produces are appropriately called AGEs (Advanced Glycation Endproducts). In conclusion, the buffering function of carbohydrates is very limited, while they seem to play a larger role in causing long-term damage than in repairing it. We may conclude that with respect to a conventional diet a wise anti-aging strategy is to reduce carbohydrate intake, while increasing protein intake, and making sure that there is a sufficient supply of essential fatty acids (De Vany, 2010; Sisson, 2009).

Another important buffer is formed by antioxidants. Antioxidants neutralize the free radicals formed by external causes (sunlight, toxins) as well by internal processes (energy production, inflammation) by donating them their excess electrons. Otherwise, the free radicals would "steal" electrons from other molecules in the body, thus damaging them. Free radicals are considered to be responsible for a major part of the accumulating deterioration that characterizes aging (Ames, Shigenaga, & Hagen, 1993; Holloszy, 1998; Slemmer, Shacka, Sweeney, & Weber, 2008). The body is capable of manufacturing its own antioxidants (e.g. superoxide dismutase or glutathione peroxidase) in case of need—an active form of regulation. However, it is easier and safer for it to rely on the widely available antioxidants from food (e.g. vitamin C, vitamin E, carotenoids, flavonoids), which circulate through the blood or which accumulate in particular organs, tissues or parts of the cell, to simply absorb any free radicals that might appear there—a buffering form of regulation. Therefore, it is a good strategy to ensure that there is always an ample reserve available of a wide variety of powerful antioxidants. This can be achieved by eating a diet high in antioxidant sources, such as berries, nuts, colorful fruits and vegetables, spices, dark chocolate, tea, coffee and red wine, and if necessary by taking a broad spectrum of antioxidant supplements.

Methyl donors are another example of nutrients with a buffering function. Molecules like choline, betaine, folate and several B-vitamins have attached methyl groups that they can donate to other molecules during critical metabolic reactions, such as DNA methylation. A scarcity of methyl donors will reduce these reactions, resulting in defects that promote aging (Anderson, Sant, & Dolinoy, 2012; Liu, Wylie, Andrews, & Tollefsbol, 2003)

A similar reasoning applies to a variety of further micronutrients (e.g. minerals, vitamins, co-factors) which the body needs to perform various functions, but which are sometimes in short supply. In such cases, according to the triage theory (Ames, 2006), the body will use the available nutrients only for the most critical functions, while neglecting long-term maintenance processes. A recurrent undersupply of such nutrients therefore accelerates aging, while a dependably plentiful supply is likely to stave it off.

A different way to extend resource buffers for the organism is rest and sleep. Most resource stocks, such as muscle or fat tissue, simply need time to (re)build. A well-investigated example of a life-extending resource produced during sleep is the antioxidant hormone melatonin (Pandi-Perumal et al., 2012). If resources are persistently depleted by reactions to disturbances, they will never reach their optimal level. Therefore, an essential part of any anti-aging lifestyle is to provide plenty of recovery time in between challenges.

Extending capabilities

The third fundamental anti-aging strategy is suggested by the $H(A)$ term in the law of requisite variety, which denotes the range or variety of actions the organism is capable of. The more actions the organism can perform, the more disturbances it can compensate, and therefore the smaller the probability that it would suffer irreparable damage. The range of actions has both a quantitative aspect (power) and a qualitative one (diversity).

Most obviously, the maximum range can be extended by increasing the maximum power or output of the physiological systems responsible for producing actions. For example, stronger muscles can neutralize larger forces or impacts on the body. A fall is a common cause of death in old people who have become weak and fragile. Higher muscular strength reduces both the probability of falling, as it helps maintain the balance, and the probability of injury, as it allows the falling person to counteract the effects of the blow. If we add the buffering role of the proteins stored in muscle mass, then it should not surprise us that lean body mass, muscle size and muscular strength have a strong negative correlation with overall mortality for older people (De Vany, 2010; Newman, Kupelian, Visser, & Simonsick, 2006).

More generally, the different components underlying physical fitness, such as lung capacity (Schunemann, Dorn, Grant, Winkelstein, & Trevisan, 2000; Sears, 2010), heart volume, flexibility of joints, solidity of bones, and flexibility of blood vessels, all contribute to the body's ability to counteract strong disturbances, and are as such good predictors of survival rates, especially for older people. For example, a person with a higher lung capacity is not only less likely to get out of breath during serious exertion, and less likely to die from drowning, asphyxiation or strangulation, but less likely to die when part of that lung capacity is compromised, e.g. because of pneumonia—another common cause of death among aging people.

While in general more difficult to observe, it is likely that physiological ranges of reaction to various internal disturbances play an equally crucial role. A liver that is capable of producing a large amount of detoxifying enzymes is less likely to let the organism die because of acute or chronic poisoning. Cells that are "insulin-sensitive", meaning that they have a large capacity for absorbing excess glucose as signaled by circulating insulin, will protect the organism much better against glycation than cells that have become "insulin-resistant"—the hallmark of diabetes and the metabolic syndrome.

Capability also has a qualitative aspect: the variety or diversity of different actions that the organism is able to perform. It is not sufficient to have great muscle strength if you cannot perform the specific movements that are necessary to break your fall, to ward off your attacker, or to get yourself out of the freezing water. The immune system further illustrates the principle: the larger the variety of antibodies available to neutralize parasites, cancer cells and other threats, the smaller the probability that some disease would overwhelm the body's defenses and thus cause you to die. While the diversity of possible reactions is more difficult to measure than their quantitative range, the law of requisite variety admonishes us to do everything we can to increase that variety. We will discuss effective strategies to achieve this goal later.

Increasing knowledge

Closely related to the variety of actions is the knowledge necessary to decide which action $A$ is appropriate for a given disturbance $D$, as expressed by the $H(A|D)$ term in our fundamental inequality (1). The smaller the knowledge, the larger the uncertainty $H(A|D)$, and therefore the higher the probability that the performed action would be inadequate. Such knowledge is expressed most simply by probabilistic *condition-action rules* (Heylighen & Joslyn, 2003):

IF the disturbance is $D$ (condition), THEN perform the action $A$, with a probability $P$

This can be represented as a weighted link

$D \rightarrow A$, with a weight $0 \leq P \leq 1$.

The bulk of that knowledge is stored in our DNA, the molecular memory that carries the wisdom accumulated during biological evolution. Specific disturbances *(D)* are translated by the organism into specific chemical signals that activate specific genes. These in turn respond by producing enzymes that set in motion the appropriate counteractions *A*.

Knowledge is also stored in our nervous system, where it is learned through experience. The basic mechanism is *reinforcement*: actions that produced benefit are more likely to be performed in the same circumstances later. In other words, success or reward strengthens the link $D \rightarrow A$ by increasing $P$. Failure, on the other hand, weakens that link; $P$ diminishes if the result of $A$ was negative rather than positive. Such reinforcement learning means that the organism becomes more likely to perform appropriate actions, and less likely to perform inappropriate ones. It in general reduces $H(A|D)$, by bringing the conditional probabilities closer to either 1 or 0. Thus, it increases the organism's ability to effectively fight off various dangers and sources of damage.

A final source of knowledge is provided by education, science and culture, i.e. by the social systems that create and disseminate new insights. If you are not able to observe for yourself what the best counteraction *A* is to a given problem *D*, your doctor, your fitness instructor, or your encyclopedia may be able to tell you. As research uncovers ever more health-related knowledge, and as information technology makes that knowledge ever more easily accessible, our stores of requisite knowledge increase very rapidly, opening up possibilities for radical life extension that we will discuss in more detail later.

One problem with knowledge is that it tends to get lost through a generalized process of forgetting. Neural connections that have not been used over a prolonged period tend to weaken, thus increasing the organism's uncertainty about how to respond to a particular condition. Both learned and inherited knowledge moreover decay through the irreversible changes that we discussed when we noted that dynamic systems tend to lose the memory of their initial states. Each time DNA is copied during the process of cell division, there is a non-zero probability that an error (mutation) is introduced that diminishes the DNA's effectiveness in initiating appropriate actions. Even static DNA can accumulate mutations through various chemical disturbances, such as attacks by free radicals. And the brain can lose its connections through strokes, beta-amyloid accumulations, misfolded prions and other causes of dementia.

In conclusion, the suggested anti-aging strategy is to maximize all forms of knowledge acquisition and learning that increase our ability to prevent or repair damage, and to minimize the factors that tend to erase memory structures.

**Challenging the organism**

We now come to a general life-extension and rejuvenation strategy aimed at extending the body's capabilities for active regulation, by increasing both the intensity and variety of actions, and their adequate targeting of disturbances (knowledge). As such, this strategy should combat the descent into rigidity and loss of systemic memory that is the fundamental cause of aging. The principle is to subject the organism to a variety of challenges of different intensities. Challenges are problems or opportunities that incite action (Heylighen, 2012a, 2012b), but that are not so difficult to overcome that they threaten to produce real losses (Blascovich & Mendes, 2000). Their function is to stir the organism into more vigorous building and repair activity than it would have performed without the challenge.

The use of challenges for strengthening body and mind has an old and venerable tradition, as exemplified by traditional methods of education and of physical, military, and character-building training. Challenges have been shown to build resilience at both the physiological (Gems & Partridge, 2008) and psychological levels (Neill & Dias, 2001; Seery, Holman, & Silver, 2010). The underlying principle is commonly expressed as "that what does not kill us makes us stronger" (Gems & Partridge, 2008; Seery et al., 2010).

The importance of challenges for health and life extension has been advocated in particular by two recent schools of thought, defending respectively the value of a *paleolithic lifestyle* and of *hormesis*. Proponents of the "paleo" approach (Carrera-Bastos, Fontes-Villalba, O'Keefe, Lindeberg, & Cordain, 2011; De Vany, 2010; Durant, 2013; Sisson, 2009) argue that our genes are adapted to the hunter-gatherer lifestyle of our paleolithic ancestors, which—in contrast to our modern, sedentary lifestyle—was characterized by much more varied and intense physical challenges, such as contact with micro-organisms and parasites, exposure to the elements, physical effort, and diverse and irregular food intake. The mismatch between our present lifestyle and the one that our genes are adapted to explains a host of life-shortening "diseases of civilization" (Carrera-Bastos et al., 2011), including CHD, metabolic syndrome, auto-immune diseases, cancer, obesity and diabetes. *Hormesis* (Calabrese & Baldwin, 2002) is the observation that a small amount of certain stressors ("hormetins"), such as heat, radiation or certain toxins, actually benefits the organism, apparently by stimulating repair mechanisms. Therefore, several authors have suggested that controlled hormesis could extend life (Gems & Partridge, 2008; Kyriazis, 2010; Rattan, 2008).

Let us first try to understand from general evolutionary and cybernetic principles why challenges appear so effective for combating aging.

Why compensatory action tends to be less than optimal

The organism has been programmed by natural selection to follow a conservative strategy of not consuming more resources than necessary. The reason is that in natural circumstances the amount of necessary and available resources tends to be unpredictable: the organism does not know how long or how intense a disturbance will be, and how much energy or food can be found to restore the reserves consumed while fighting off the disturbance. For example, a hunter cannot know in advance how much effort it will cost to catch a prey animal. Similarly, when the organism tries to heal a minor wound, it does not know exactly how much time or how many resources it will need. Therefore, the energy invested in counteracting a disturbance that is not immediately threatening will in general be less than the energy that would optimally solve the problem. The organism simply cannot afford to put all its resources into tackling a specific problem, because even if it was certain that this action would solve the problem (which it is not), other problems may well appear that require these resources more urgently.

This is a complement to the short-term vs. long-term trade-off that we discussed earlier. The organism not only must balance urgent needs with those that only manifest themselves in old age, it also must trade off present problems against potential, unknown ones, which could strike at any moment. The only reliable method is to maintain strategic reserves: do not spend all your energy running to catch that deer, because you may need it to fight off the tiger that is waiting behind that bush to attack you!

The practical consequence is that our disturbance-combating reactions are in general *suboptimal*: they will not use all the available resources to prevent or repair damage. This counterintuitive effect can be illustrated by an observation from sports science: competitors in a race always go fastest when they are close to the finish—even though that is the moment they should be most exhausted. A complex systems model of fatigue (Lambert, Gibson, & Noakes, 2005) proposes that their body has been saving resources during the race as reserves for unforeseen contingencies by producing a feeling of fatigue that prevents additional effort, but that the nearby end of the race tells the brain that these reserves can now be released, thus reducing the fatigue at the critical moment. On the other hand, when the finish of the race is unknown, the brain tends to be even more cautious in conserving resources, creating a more pronounced feeling of fatigue. Therefore, the body's experienced capability for effort is reduced, even though objectively the difficulty of running a given distance is the same as when the finish is in view.

While for an organism in the wild it is wise to maintain reserves for unforeseen contingencies, this principle applies much less to us, people of the industrial age: we know that we can quickly restock with any necessary resources, and we can also be pretty confident that we will not suddenly have to run from a tiger, escape from a fire, or try to survive in ice-cold water. This means that it would be good if we could override this in-built safety mechanism, and force the organism to throw more resources at the problem. That would provide for a more extensive repair of damage, and an expansion of our capabilities for action, thus counteracting any reduction of the homeodynamic space.

## Why challenges need to be sudden and intense

The trick for achieving such an override is to challenge the organism with a short, but intense and unusual disturbance so that it is forced to react by mobilizing extra resources. If this is supported by an abundant reserve of such resources, so that there is little danger of running out, the organism is more likely to overshoot (use more resources than necessary) than to undershoot (use less).

The reason for the overshoot is the uncertainty about the seriousness of the challenge: better hit an attacker hard enough the first time so that he won't come back, than run the risk of a second attack that may kill you. If the potential destructiveness of the challenge is not known, but the indications are that it is powerful, the safer strategy is to mobilize more than may be needed. On the other hand, for well-known, moderate challenges, the organism can estimate exactly how much it will minimally need, and save the rest. The effect of the overcompensation is that more damage will be repaired than the one caused by the disturbance, thus actually rejuvenating the organism. Moreover, the organism will have learned to prepare for more intense challenges than it was used to, and therefore to build up its capabilities for tackling similar problems in the future. This increases its range of action or "homeodynamic space".

A well-studied example of this dynamic is the effect of overheating. When the body is subjected to unusually high temperatures (e.g. in a sauna), it reacts in part by producing Heat Shock Proteins (HSPs) (Feder & Hofmann, 1999), which repair damage caused by heat and other stresses to molecular structures. These HSPs diffuse throughout the body, removing or reassembling any misfolded proteins—not just those produced by the heat. Therefore, the controlled administration of heat stress can be expected to reduce the deterioration caused by aging (Feder & Hofmann, 1999). Such a life-extending effect has effectively been demonstrated in a number of model organisms and cell cultures (Rattan, 2008).

An even better known example of the capability-boosting effect of challenges is exercise. Vigorous physical exertion puts a heavy stress on the body: glucose and oxygen levels are depleted, large amounts of free radicals are produced, muscle fibers are torn, and lactic acid accumulates in tissues. These deviations from the ideal range of the essential variables must be corrected by a variety of compensatory, repair and building actions. One of the more immediate effects is that the body will generate large amounts of extra antioxidants to combat the free radicals. The accompanying overshoot is likely to clear out some free radical damage dating from before the exertion. In the longer term, the processes that generate these endogenous antioxidants are likely to be reinforced, so that they are ready for further onslaughts. The net effect is a body better prepared to deal with free radical damage. A similar effect is seen in the muscles: the torn fibers are disassembled through a local inflammation process, and then regrown in a stronger version, so that they are less likely to suffer similar damage in the future. The whole process obviously consumes a lot of extra resources, which is why the body will not go through that process unless it is forced to by a sufficiently powerful challenge. This explains why athletes need to eat more than people with a sedentary lifestyle—but that is not really a constraint in our present society.

Such a hormetic effect of physical stress has recently been promoted as an anti-aging strategy by several authors (Gems & Partridge, 2008; Kyriazis, 2010; Rattan, 2008). However, the original definition of hormesis (Calabrese & Baldwin, 2002) is a dose-dependent effect of stressors: initially, for small amounts, the effect is positive, but as the dose increases, the effect becomes increasingly negative. This seems to imply that there is a fixed, optimal amount for each specific stressor—e.g. a toxin. But the challenge interpretation says more or less the opposite: to get the maximum benefit, the stressor should be irregular, unexpected, and as intense as practicable without running the risk of irreversible damage. Indeed, the function of the challenge is to mobilize resources that the organism would normally keep in reserve during more routine disturbances. Constant, predictable levels of stress would quickly lead to adaptation, where the organism would learn just how much resources it needs to prevent any immediate dangers, and save the rest for unexpected problems.

The difference between both strategies can be illustrated by comparing two different styles of physical exercise: traditional "cardio" training and High Intensity Interval Training (HIIT, (Gibala & McGee, 2008)). The most common way people train is by regularly jogging for extended periods at moderate intensity, making sure that their heart rate stays within the aerobic training regime (around 75% of the maximum heart rate). Such a regime leads to clear cardiovascular improvements in the short term. However, in the longer term fitness tends to stagnate, and long-distance runners even tend to lose muscle mass while suffering wear and tear injuries in the joints from chronic overuse, and occasional heart-attacks during marathon sessions (Sisson, 2009).

The HIIT regime, on the other hand, requires a series of short, but very intense "sprints", where you try each time to do more than what you thought you were capable of, interspersed by periods of recovery. The positive effects are much more pronounced than with the "cardio" regime: cardiovascular fitness, lung capacity, lean muscle mass, and strength increase rapidly, for a much shorter overall training duration, and without any apparent negative effects or plateaus (Helgerud et al., 2007; Sears, 2010; Sisson, 2009).

Long-lasting challenges, like in cardio training or traditional hormesis, are in effect a drain on the organism's resources: it continuously needs to repair small levels of damage, caused e.g. by free radicals or pressure on the joints, but in order to save resources for unexpected problems, it will do this suboptimally. Moreover, it will tend to save resources on capabilities that are not fully used, such as muscle mass in the case of slow-moving long-distance runners. Therefore, after the initial boost they give to the body's capabilities, the positive effect of such challenges will erode and perhaps even turn negative. Sudden, brief challenges of gradually increasing intensity, on the other hand, tend to produce overshooting reactions in anticipation of future, stronger challenges, while leaving plenty of time in between to fully repair the damage and rebuild the resources.

A similar dynamic is well known for psychological stress. Our body and mind are much better prepared for acute stress (intense but brief) than for chronic stress (moderate but enduring). After the initial fright, running away from a tiger makes you feel exhausted, yet exhilarated. Enduring the on-going time pressure at work, on the other hand, merely makes you feel increasingly drained, until you break down with a depression, ulcer, or worse…

Why challenges need to be as diverse as possible

The intensity of challenges increases the quantitative range of the body's capability for action. However, the law of requisite variety reminds us to also increase their qualitative range, or diversity. This can be achieved through a greater variety of challenges: the more differentiated or diverse the disturbances that the organism is subjected to, the more diverse the abilities it will need to develop in order to cope with them.

Typical sports training, like running, cycling, or lifting weights, focuses on one or a few types of movement. *Crossfit* is a recently popular approach that tries to broaden the range of physical skills that are trained, in order to involve all the muscles and functions of the body. Yet, its routines remain largely fixed (Durant, 2013). More complex and unpredictable movement abilities are developed in certain martial arts, such as judo, and certain team sports, such as basketball, where participants need to surprise their opponents with unexpected moves in order to break through their defenses. Similar requirements characterize sports—such as rock climbing, wild water rafting, or hunting—that are performed in typically unpredictable natural environments. A particularly inspiring new training method is *MovNat*, developed by Erwan Le Corre and inspired by the Paleolithic lifestyle, in order to teach people to "move naturally" and "move in nature", while developing the functional abilities that the body would need to survive in a variety of challenging situations (Durant, 2013).

There is as yet no evidence that more varied training would be life extending, but a strong indication can be found in research on *environmental enrichment* (EE). Animals used for experiments, such as mice and rats, are normally kept in extremely simple and predictable environments: an empty cage, where they receive always the same food, at the same moments. An enriched environment provides a variety of stimuli or challenges, such as toys, exercise wheels, ladders, and a larger number of individuals to interact with, spread over a larger space. Animals living in enriched environments do better on a variety of measures: EE stimulates the creation of new neurons and synapses, rejuvenates brain function, strengthens the immune system, reduces the incidence of cancer, and generally extends life (Arranz et al., 2010; Fabel et al., 2009; Mora, Segovia, & del Arco, 2007; van Praag, Kempermann, & Gage, 2000). This happens in part because EE (together with exercise) incites the production of BDNF (Brain-Derived Neurotrophic Growth Factor), an endogenous molecule that plays a role similar to HGH (Human Growth Hormone) in stimulating the (re)building of tissues that were particularly challenged—with HGH focusing on muscles and BDNF on neurons.

Environmental enrichment appears to extend both physical and cognitive abilities: the organism increases both its variety of actions (the $H(A)$ term) and its knowledge about how to use these actions most effectively (the $-H(A|D)$ term). The knowledge effect is due to the fact that the environmental challenges, while complex, are not random: repeatedly tackling the same kind of problem (e.g. climbing a ladder) will differentially reinforce adequate moves and weaken inadequate ones. Thus, the organism learns to perform a coordinated pattern of actions, which is stored in the highly non-random pattern of connections between its neurons and synapses. It is the need for such complex knowledge, as signaled by BDNF, that forces the organism to invest resources in neural growth and rewiring.

At a more practical level, the EE strategy for rejuvenation can be implemented by regularly confronting yourself with new problems, experiences and perspectives, ideally together with physical exercise (Fabel et al., 2009). An effective combination of such physical, cognitive and emotional stimulation can be found in adventure-style activities (Bowen & Neill, 2013; Neill & Dias, 2001; Neill, 2008), where people climb a mountain, explore a wilderness, or travel across a remote region, while having to improvise solutions to various unforeseen challenges (Heylighen, 2012a). The traditional advice to solve crossword puzzles in order to stave off brain degeneration seems rather ineffectual in comparison, because the strategies used to solve these puzzles remain essentially the same, while mobilizing only very specific intellectual skills, and no motor skills. The same applies to a lesser degree to various "brain training" computer games that purport to increase capacities such as reaction speed, working memory, or focus: while they may strengthen some existing abilities, they do not really call forth a reorganization of your neural networks. Better suggestions would be to take up a new sport or avocation, to learn a new language, to get to know new people, to visit other countries, to explore new disciplines and art forms, and to broaden or change the manner and focus of your professional activities.

However, as with physical challenges, the rule remains that it is better to have relatively short, intense challenges interspersed with periods of recovery long enough for the organism to build up its capabilities and replenish its resources. Mental resources too get exhausted by chronic stress—as exemplified by the all-too-common phenomenon of professional burnout. Therefore, the typical life of an expatriate, who lives in one country doing one job, and then relocates to another, having to adapt to a completely different linguistic, social, cultural, professional and physical environment, just in order to start the whole process again a year or so later, appears to be more draining than stimulating.

Maximizing variety does not just apply to movement and to cognition: our diet too should be as varied as possible. This ensures both that we accumulate a wide range of protective, buffering nutrients, such as antioxidants, and that we get exposed to a wide variety of toxins to challenge the organism's detoxification capabilities. Several authors suggest that the disease-preventing and life-extending effects of many vegetables, such as those from the cabbage family, and spices, such as turmeric and pepper, are actually due to the hormetic effect of the toxins they contain in order to ward off predation (Kyriazis, 2010; Rattan, 2008).

A similar logic applies to exposure to micro-organisms and parasites: disease-producing organisms will teach the immune system to effectively combat them; symbiotic ones will increase the diversity (and thus the resilience and buffering capability) of the bacterial ecosystem that lives in our intestines and skin. The "hygiene hypothesis" attributes the present increase in allergies and auto-immune diseases to a reduction in the number of pathogens that challenge people's immune system (Okada, Kuhn, Feillet, & Bach, 2010; Yazdanbakhsh, Kremsner, & van Ree, 2002). Therefore, frequent washing, disinfection, and avoidance of contact with "dirt" are more likely to weaken than to protect the organism.

Why random disturbances are good for you

Building knowledge requires complex, consistent challenges that elicit intelligent reactions. But completely unstructured, random challenges also have a rejuvenating effect. This can be understood from another cybernetic principle, which has been called "order from noise" (Heylighen, 2001; Von Foerster, 1960), "order through fluctuations", or "order out of chaos" (Prigogine & Stengers, 1984).

A simple illustration is how you can pack small pieces (e.g. dry beans, nails, tea leaves, sand grains…) more densely in a container. You first fill the container (e.g. a jam pot) to the brim, close it, and then chaotically shake it and hit it, starting with more violent movements and slowly easing off into more gentle ones until you stop. You will notice that the initially full container now offers plenty of empty space at the top, allowing you to add more pieces. The principle is that the initial arrangement of pieces is far from optimal: there are quite some empty spaces left between the beans or nails. But this suboptimal arrangement is unstable, because gravity incites the pieces to move as closely as possible to the bottom of the container: if a bean would be positioned on top of an empty space it would drop down to fill it. But to achieve that, it needs to move into the right position at the right moment, when the space opens up. This is unlikely to happen for all spaces when you initially pour the beans in the container. However, when the beans are shaken, they get the chance to explore many different arrangements. The unstable ones will be destroyed by the next shake, but the most stable ones, where the beans are most densely packed close to the bottom, will be maintained. In this way, all the unstable, fragile configurations are selectively eliminated by the random movements. The end result is a much more robust arrangement in which the components are nearly optimally aligned.

A similar example can be found in the process of *annealing* that is used to make metal or glass stronger and more flexible. The metal is repeatedly subjected to strong heating, which increases the random movements of the atoms, and then allowed to cool down slowly. Thus, the atoms get the opportunity of trying out many different configurations in order to eventually settle down into the most stable one. This is normally a regular crystal lattice, without any defects or "empty spaces"—i.e. weak spots where the metal is likely to break when put under stress. Thus, "order" is enhanced by the addition of random noise or chaos to the system. Taleb (Taleb, 2012) calls this property of gaining strength from disorder *antifragility*.

Order from noise or antifragility can be understood as an application of the evolutionary mechanism of blind variation and selective retention (Heylighen, 1992) to the internal structures that make up a system. The more destabilizing variation these structures are subjected to, the more of the weak, fragile or unfit arrangements will be eliminated, and replaced by stronger ones—until only the fittest remain. In economic evolution, this phenomenon is known as *creative destruction* (Metcalfe, 1998). In personality development, it has been called *positive disintegration* (Dabrowski, 1967; Laycraft, 2009).

The variation is ideally random or non-directed, because targeted variation will only eliminate the problems it is directed at, while fragility can emerge in any part or at any level of the system organization. It results from what we called the "descent into rigidity", the automatic process that tends to replace flexible, adapted structures with rigid, non-functional ones—unless challenged—, and which is the main driver of aging.

Weight training is an example of an external variation that creates a directed selective pressure on the body: subjecting muscles to strong forces will tend to break the weakest fibers in those muscles, stimulating the organism to produce stronger ones that can survive the force next time. But depending on the direction of the force, only certain types of fibers will undergo this rejuvenating effect. The effect can be broadened by adding an essentially random component to the challenge, such as a vibration independent of the direction of the force. This impels additional fibers to kick in, in order to neutralize disbalances, sideways movements, and other deviations created by the rhythmic shaking. This is the mechanism behind *vibration training*, in which people perform exercises on a plate that vibrates with different frequencies and amplitudes in order to enhance the training effect (Luo, McNamara, & Moran, 2005). In addition, we should expect vibration to "shake up" joints, bones, internal organs and tissues in a gentle but wide-ranging manner, selectively destroying brittle configurations and thus forcing the organism to rebuild them more flexibly. The technique is particularly promising for enhancing bone mass, strength, speed and other capabilities in aging people (Merriman & Jackson, 2009; Roelants, Delecluse, & Verschueren, 2004).

Another example of a wide-ranging, undirected source of variation is the circulation of blood and lymph throughout the body. While this is an intrinsically dynamic, flow-like process, circulation too is subject to the phenomenon of a spontaneous "descent into rigidity" in which the range of flow is gradually restricted—as exemplified by narrowing or formation of obstructions in vessels, reduced flexibility of vessels walls, and the reduced irrigation of certain organs and tissues. To counteract this tendency, from time to time we need a powerful injection of "noise" into the circulatory system, by forcing it to flow with speeds, pressures and directions that is not used to, in order to "flush out" any obstructions, accumulated toxins or debris.

A simple technique, which has been applied successfully since ages to relieve fatigue, pain and inflammation, is to alternately heat and cool (part of) the body—e.g. with hot and cold compresses, or sauna sessions followed by immersion into a freezing lake. The heating increases blood flow through the affected body parts, while the cooling constricts the vessels, pushing out liquid. In succession, they are likely to clear out any stagnating "waste materials", such as toxins (Crinnion, 2007), damaged cells, or free radicals produced by inflammation, while flooding the tissues with fresh oxygen, antioxidants and nutrients, and training the vessels to become more flexible. Regularly subjecting the body to large temperature changes generally improves circulation while stimulating the immune system (Brenner et al., 1999; Durant, 2013).

Intense exercise achieves to some degree a similar effect by forcing the heart to pump blood much more powerfully through the vessels. A further technique to help "shake up" the circulatory system is yoga. The different yoga postures ("asanas") stretch or compress different tissues and organs, thus stimulating their flushing. The upside-down postures, where the head is on the floor and the legs up in the air, moreover inverse the normal effect of gravity on the blood flow, creating a flush of blood to the brain, while draining the lower extremities. In addition, yoga increases the flexibility and strength of joints, tendons and muscles by subjecting them to unusual, but controlled stresses, while relaxing and focusing the mind. Here too there is evidence of a rejuvenating effect (Oken et al., 2006; Wang, 2009).

A final example of a "non-directed" disturbance is fasting. There is a lot of experimental evidence for the life-extending effects of *caloric restriction* (CR, reducing the amount of food consumed to below the hunger level) on short-living animals. However, the most advanced trial with rhesus monkeys seems to indicate that these benefits may not extend to the longer living primates (Mattison et al., 2012; Maxmen, 2012)—at least if both calorically restricted and control groups are fed a varied and nutritious diet. The benefits of caloric restriction have been attributed to hormesis: a limited amount of a bad thing (starvation) may actually have positive effects, by boosting the body's defense systems (Kyriazis, 2010; Rattan, 2008).

However, the challenge model proposes that the positive effect of such stresses is greater for short, intense challenges (not eating at all for one or a few days) than for continuing, moderate challenges (eating everyday some 30% less than normal, for the rest of your life). The former strategy, where you regularly pass a day without food, but eat as much as you want otherwise, has been called *intermittent fasting* (Anson, Guo, Cabo, & Iyun, 2003; Brown, Mosley, & Aldred, 2013; Varady & Hellerstein, 2007). While there is as yet much less research on this topic, the indications are that it has benefits similar to CR (improved metabolic markers, reduction of cancer, diabetes and cardiovascular disease, general rejuvenation…), but without the shortcomings of CR (chronic feelings of hunger, coldness, lack of energy, loss of muscle mass…). These shortcomings actually seem serious enough to *decrease* the life expectancy of calorically restricted animals outside of the strictly controlled laboratory conditions in which CR has been tested, because they reduce the organism's capability to cope with common challenges, such as infections, injuries, and temperature extremes (Adler & Bonduriansky, 2014).

The most important underlying rejuvenation mechanism appears to be that fasting stimulates the process of *autophagy* ("self-eating"), in which unnecessary cellular components (proteins, organelles…) are broken down and recycled (Bergamini, Cavallini, Donati, & Gori, 2007; Morselli et al., 2010; Rubinsztein, Codogno, & Levine, 2012). The less food the organism gets, the more it will be driven to digest internal components in order to extract the nutrients necessary for normal functioning. Physical exercise appears to have a similar effect (He, Sumpter, Jr., & Levine, 2012), perhaps because it acutely increases the demand for nutrients. Autophagy seems to selectively consume damaged components, thus clearing out dysfunctional material. When nutrients become available again, these components will be rebuilt, thus rejuvenating the machinery of the cell. But even if autophagy were not targeted specifically at dysfunctional components, it would still reduce the proportion of such components, and thus counteract aging. While there is as yet no evidence for this, it seems likely that a similar logic applies to *apoptosis*, the process that breaks down and recycles complete cells (Adler & Bondurianski, 2014). This would explain the cancer-preventing properties of CR and exercise, because precancerous cells would be the first targets for apoptosis.

**The buffering-challenging strategy: summary**

The principles of cybernetics have provided us with a foundation for understanding both the processes of aging and the processes that counteract them. The extended law of requisite variety formulates in the most general terms what a system needs to survive: a limited range of disturbances that could endanger it, a large enough variety of actions through which it could neutralize these disturbances, sufficient functional knowledge about which action to perform in which circumstances, and a large enough reserve of resources to buffer unpredictable fluctuations. Ashby's principle of self-organization tells us that dynamic systems—including organisms—tend to spontaneously converge onto a smaller, more constrained part of their state space. This reduces their freedom to respond to various challenges, i.e. their variety of action. It results in an increasingly rigid and brittle state, while eroding the knowledge necessary to apply actions effectively. In principle, the effects of this spontaneous process of deterioration can be counteracted in the same way as ordinary disturbances, by using the appropriate actions, knowledge, and reserves to repair or prevent any damage to functional structures.

However, while natural selection has equipped us with control mechanisms that cope very effectively with short-term damage, these mechanisms are much less effective in the long run. The reason is that no control mechanism is able to neutralize *all* disturbances. Given the relative costs and benefits of different actions, trade-offs have to be made and priorities chosen. Generally, intense and immediate challenges take priority over slow and far-away processes of deterioration. This means that the organism will not invest all its resources in combating aging, but in making sure to keep a critical reserve for dealing with direct challenges, actual or potential. The result is a gradual accumulation of damage that reduces the functional range of activity. As the variety and functionality of action decreases, the range of deviations from the ideal state extends, increasing the amount of further damage. Thus, deterioration is a self-amplifying process, characterized by accelerating fragility and eventually mortality.

Such a strategy that withholds resources from long-term processes in order to save them for short-term disturbances makes sense in a natural environment, where resources are limited and unreliable, and lethal dangers are common. However, in our present industrialized society, it is no longer useful: we have all the resources we need, while acutely life-threatening situations have become very rare. Therefore, if we want to maximize our life span, we would do well to override this strategy that natural selection has programmed into our genes. While most researchers in the aging field try to manipulate genetic programs via specific metabolic pathways, I have proposed a general counterstrategy that co-opts our evolved program of reactions to redirect resources towards damage repair. Since repair processes depend on the level of challenge and the level of resources or buffers, we can manipulate them simply by controlling these levels.

First, we should make sure that the body has all the resources it needs. This applies in particular to micronutrients, such as vitamins, amino acids, and minerals, which tend to be less plentiful in industrially processed food than in the "wild" foods eaten by our Paleolithic ancestors (De Vany, 2010; Sisson, 2009), and which are subjected to a triage process that allocates them with priority to short-term processes (Ames, 2006). Moreover, many of these micronutrients, such as antioxidants and methyl donors, play a direct buffering role by absorbing local disturbances, thus making costly repair processes unnecessary.

Second, we should subject our organism to the kind of challenges that elicit powerful responses and thus effectively mobilize these resources. The response to an external stressor will typically have two components (Demirovic & Rattan, 2013): (1) an immediate action powerful enough to *prevent* (feedforward control) or *repair* (feedback control) any serious damage; (2) a delayed reinforcement of the mechanisms responsible for producing such actions, so that they would be better prepared for similar challenges in the future. When the challenge is sufficiently intense and unusual, the immediate response will tend to *overcompensate*, mobilizing more resources than strictly necessary. This normally clears up more damage than induced by the stressor, thus effectively rejuvenating the organism. The delayed response too will rejuvenate the organism, by extending its range of action or homeodynamic space, and thus reducing rigidity and fragility.

Effective challenges are ideally brief, since a prolonged challenge tends to deplete resources, while inducing habituation that decreases the response back to a suboptimal level. Life-extending challenges are also as diverse as possible, so as to elicit a maximum range of actions and capabilities for action, as demanded by the law of requisite variety. There are two strategies for increasing diversity: targeted and random. A targeted strategy is directed at a specific problem domain, trying to increase capabilities and knowledge for acting towards challenges in that domain. Increasing diversity can be achieved by targeting complex, unpredictable domains, where many, interrelated things can happen, each demanding its own finely coordinated combination of actions. It can also be achieved simply by expanding the range of domains, i.e. by constantly exploring new types of challenges. The random strategy assumes that we will never know *all* the domains in which something can go wrong. Therefore, it is good to seek out challenges that are as wide-ranging and unspecific as possible in their effects—an equivalent of the "random shaking" that flushes out fragile arrangements, wherever in the system these weak spots may hide. This is an application of the order from noise principle, or the idea of creative destruction.

For a practical application of the strategy with suggested interventions, I refer to Table 1.

| Strategies | Types | Examples |
|---|---|---|
| *Avoiding dangers* | Causes of irreversible damage | Avoidance of smoking, high-glycemic diet, potentially lethal dangers, heavy metal accumulation, chronic inflammation, chronic stress… |
| *Building up buffers* | Antioxidants | Cofactor Q-10, grape seed extract, tea, coffee, carotenoids, flavonoids, Acetyl-L-Carnitine, Alpha Lipoic Acid, resveratrol, quercetin, milk thistle… |
| | Methyl donors | Choline, betaine, folate, TMG… |
| | Vitamins | A, B, C, D, E, K… |
| | Minerals | Se, Mg, Ca, K, Fe, I, Mn, Zn, … |
| | Amino Acids | Arginine, cysteine, methionine, glutamine, leucine, valine, … |
| | Essential Fatty Acids | DHA, EPA, linoleic acid, … |
| | Rest & sleep | |
| *Subjecting the organism to challenges* | External | Heat, cold, UV-light, aerobic exertion, anaerobic exercise, stretching, compression, vibration, exposure, … |
| | Internal | Infections, parasites, wounds, radiation … |
| | Diet | Fasting, feasting, plant toxins, alcohol, drugs, spices, … |
| | Mental | Environmental enrichment, adventure, study, research, new avocations, acute psychological stress, … |

**Table 1**: a list of anti-aging interventions suggested by the buffering-challenging strategy, with typical examples. Note that some examples fit in more than one category (e.g. spices and flavonoids can function both as antioxidants and as mild toxins stimulating repair processes). The dangers to be avoided produce long-term, cumulative damage; the challenges to be sought produce brief disturbances that boost repair mechanisms.

**Implications for Radical Life Extension**

While this general, "cybernetic" approach towards life extension and rejuvenation is new, most of its components have already been proposed and investigated by different authors, in particular from the "paleo" (De Vany, 2010; Durant, 2013; Sisson, 2009) and "hormetic" (Gems & Partridge, 2008; Kyriazis, 2010; Rattan, 2008) schools. As such, there is quite a bit of evidence showing that recommended interventions, such as fasting, vibration, interval training, plant-based antioxidant supplementation, micronutrients, environmental enrichment, and exposure to heat and cold have rejuvenating and even life extending effects, at least in model animals. But there are as yet no empirical data about the effects of a systematic, combined strategy like the one I am advocating here. Cybernetic reasoning suggests that the different interventions would be mutually reinforcing, thus producing an overall benefit that is more than the sum of its parts. This can be understood by examining the positive feedback involved in aging.

By the time people reach their forties or fifties, their body composition, health markers, physical condition, and energy level have typically deteriorated significantly relative to their best years. As a result, they are less inclined to take on various physical challenges, and especially novel ones. What this leads to becomes clear in people who are one or two decades older. As their capabilities no longer are challenged, they become weaker—an illustration of the *use it or lose it* adage. As they become weaker, they avoid challenges even more. As a result, they become weaker still, and avoid an even broader range of challenges, …, until they are so weak that the next, unexpected challenge kills them. According to the cybernetic model I have proposed, it is this vicious cycle of fragility engendering more fragility that explains the accelerating rate of mortality that characterizes aging.

The buffering-challenging strategy, on the other hand, reverses the direction of this positive feedback, with more difficult challenges producing higher capabilities, which in turn stimulate people to take on even more difficult challenges. The effect is a virtuous cycle of robustness or adaptivity engendering more adaptivity. This should at least significantly postpone the aging process. Both the evidence I have summarized here and the personal experience of people who have applied similar programs (e.g. (De Vany, 2010; Sisson, 2009) and also my own experience) seems to confirm this.

But could aging be postponed forever? All the experience of anti-aging interventions until now seems to say that it *cannot*: no person or experimental animal has ever been able to live indefinitely. Yet, there are reasons to remain optimistic. First, some types of organisms, such as bacteria, certain plants, and hydras, do not appear to age: they seem able to stave off death indefinitely by constantly renewing the organism via growth, cell division or budding. Second, both evolutionary theory and empirical evidence suggest that for the other organisms at some very late stage in life *aging stops* (Mueller et al., 2011)—in the sense that mortality rates do not increase any further. Unfortunately, at that stage mortality rates are already so high that continuing survival is extremely unlikely. But if we would be able to sufficiently postpone the growth of fragility that drives up mortality rates, we might hope to reach that "old age plateau" in relatively good condition. At that point, continuing survival would merely require a continuing success in counteracting any disturbances that may appear, without further fear for an irreversible deterioration. And that would in principle allow indefinite life extension!

Let us examine in more detail what this means. In an infinitely extended universe, the variety of disturbances that could potentially hit an organism is infinite. Given enough time, any event or combination of events anywhere in the universe may eventually have a causal impact on a long-living organism, and thus disturb it in some novel manner. For example, if you survive for another five billion years, you are likely to be killed by the Sun turning into a red giant and swallowing the Earth. However, you are much more likely to have died earlier because of some complex combination of molecular-level events unforeseen by your genetic programming (or by your doctor), whose emergent effects disrupt some vital component of your survival machinery. According to the law of requisite variety, being able to control an infinite variety of disturbances would require an infinite repertoire of potential counteractions, and an infinite amount of knowledge about when and how to perform these actions. Since it is impossible to store such an infinite regulatory capacity within a finite organism, this appears to preclude any form of immortality.

However, in mathematics infinity is defined as a *virtual limit*—an extrapolation to the end of an iteration that does not end. In actuality, all phenomena are finite, and therefore no organism will ever have experienced more than a finite number of disturbances—albeit a variety that grows without limit. Therefore, the requirement for unlimited life extension is that the variety of actions and of knowledge would grow at least as fast as the variety of actually experienced disturbances. To continue living, it suffices that for every novel disturbance that appears you would have learned to perform an effective counteraction *before* it induces irreparable damage. Thus, you would be able to postpone death *indefinitely*—i.e. not *infinitely*, but without a priori defined limit.

The infinite variety of potential disturbances imposes a clear limitation, though: there are no miracle solutions to the problem of aging. We will not achieve unlimited life extension by the application of any specific drug, intervention, or technology, because no single application has the requisite variety to pre-empt *all* potential disturbances. Evolutionary biologists have by now clearly demonstrated that there is no specific "aging program" in the genes. Therefore, there is not a "switch" somewhere in the organism that we could simply shut off in order to live forever. Instead, theory and observations portray aging as the accumulated effect of a seemingly endless complex of interacting genetic and environmental influences (Mueller et al., 2011; Rattan, 2012; Rose et al., 2014). As yet, we have managed to only pinpoint the most obvious causes of accumulating damage—such as oxidation, glycation, telomere shortening, and DNA transcription errors. As we develop effective interventions to combat these processes of deterioration, we should expect that this advance will merely reveal further, as yet unknown sources of disturbance.

Nevertheless, if we continue growing in knowledge and capabilities quickly enough, we should in principle be able to keep up with this relentless onslaught of ever-novel challenges. Science and technology are advancing ever more quickly, and so are our capabilities to prevent or repair specific types of damage to our health. The result is that life expectancy in most countries is increasing at a rate of about 1 year extra for every 3 years. This trend is part of the "demographic transition" in which populations that develop—economically, socially and technologically—undergo a shift from high mortality and high fertility to long life and slow reproduction. In the limit, this shift would result in zero mortality and zero fertility (Heylighen & Bernheim, 2004; Last, 2014). Quantitatively, the only thing needed to reach this limit of indefinite life span is to triple the rate of advance in life expectancy to one extra year for every year that passes!

At the moment, such a drastic increase does not seem realistic at the population level. However, an intelligent and motivated individual who keeps up with all the most important advances in the field may be able to stave off death long enough in order to live through to the next great anti-aging breakthroughs. This would buy her another decade or so to assimilate further advances. During this decade, further life extending breakthroughs are likely to happen, buying her another decade, and so on—forever (Kurzweil & Grossman, 2005).

To practically implement such a scenario, I suggest combining the broad cybernetic approach I have sketched with the capabilities for knowledge discovery offered by the Internet. The buffering-challenging strategy provides a broad foundation for radical life extension because it is *open-ended*:

it admonishes us to constantly try out new buffers, challenges, and actions in order to build new capabilities and learn new knowledge. Moreover, it seeks to maximize the variety and reach of order-from-noise processes, thus potentially "flushing out" problems that modern science has not even detected yet.

The Internet, on the other hand, is the perfect instrument for targeted search and exploration of existing and novel ideas, techniques and tools. Furthermore, it allows you to nearly instantaneously purchase the various supplements, medicines, and technological supports that you may need to implement your chosen intervention. Thus, it boosts both requisite knowledge and requisite variety of action. Finally, it boosts the collective intelligence of the community of researchers addressing these various, intermingling issues related to health and aging. Thus, the emerging "global brain" formed by the world community supported by the Internet accelerates scientific discovery to such a degree that it may become an automatic process (Heylighen, 2014), answering questions nearly as soon as they are asked. The combination of this endless flux of new knowledge and tools with the healthy, physical basis provided by a body that is systematically buffered and challenged appears like a solid platform for indefinite life extension.

Initially, such radical life extension will only be achievable for a select group of very intelligent, highly educated, and highly motivated people, who are capable and willing to perform the endless research and self-challenging that is necessary to implement the proposed strategy. But as happens with all innovations, the necessary skills will gradually diffuse to larger and larger groups, boosting their life expectancies as well. Hopefully, the ensuing drastic reduction in mortality will remain more or less in balance with the on-going reduction in fertility, so as to avoid too large changes in population level (Last, 2014).